\documentclass[twocolumn,english,Xlinenumbers,aps,twocolumn,groupedaddress,superscriptaddress,showpacs]{revtex4-1}
\usepackage{amsmath,amssymb,graphicx}
\usepackage{epstopdf}
\usepackage{graphicx}
\usepackage{dcolumn}
\usepackage{bm}
\usepackage{babel}
\usepackage[unicode=true,pdfusetitle,
 bookmarks=true,bookmarksnumbered=false,bookmarksopen=false,
 breaklinks=false,pdfborder={0 0 0},backref=false,colorlinks=true]
 {hyperref}
\hypersetup{citecolor=blue}

\usepackage{color}

\begin{document}

\title{Stimulated scattering in strongly coupled nanolasers induced by Rabi oscillations}

\author{Mathias Marconi}
\affiliation{Centre de Nanosciences et de Nanotechnologies, CNRS, Univ. Paris-Sud, Universit\'e Paris-Saclay, C2N-Orsay, 91405 Orsay cedex, France }

\author{Julien Javaloyes}
\affiliation{Departament de F\'isica, Universitat de les Illes Baleares, C/Valldemossa km 7.5, 07122 Mallorca, Spain }

\author{Fabrice Raineri}
\affiliation{Centre de Nanosciences et de Nanotechnologies, CNRS, Univ. Paris-Sud, Universit\'e Paris-Saclay, C2N-Orsay, 91405 Orsay cedex, France }

\author{Ariel Levenson}
\affiliation{Centre de Nanosciences et de Nanotechnologies, CNRS, Univ. Paris-Sud, Universit\'e Paris-Saclay, C2N-Orsay, 91405 Orsay cedex, France }

\author{Alejandro M. Yacomotti}
\email{Alejandro.Giacomotti@lpn.cnrs.fr}
\affiliation{Centre de Nanosciences et de Nanotechnologies, CNRS, Univ. Paris-Sud, Universit\'e Paris-Saclay, C2N-Orsay, 91405 Orsay cedex, France }

\date{\today}


\begin{abstract}
Two coupled-cavity systems, or "photonic dimers", are efficient test-beds for both fundamental optics –the realization of quantum correlated states, Josephson physics, and so forth–, and applications such as optical flip-flop memories. 
In this work we report on the first observation of nonlinear mode interaction in a photonic dimer formed by two semiconductor photonic crystal coupled nanolasers. For this, we investigate energy transfer between hybrid modes, which manifests as a switching from the blue-detuned (bonding) to the red-detuned (anti-bonding) modes. An mean-field model allows us to explain this phenomenon as stimulated scattering due to carrier population oscillations in the cavities at the Rabi frequency. Such asymmetrical mode interaction is universal in semiconductor laser photonic molecules, and unveils the origin of cross-correlation dips in the statistics of mode fluctuations.
\end{abstract}

\maketitle 

\section{Introduction}

Semiconductor coupled microcavity systems, also called photonic molecules (PMs), are attracting considerable attention as efficient test-beds for quantum and nonlinear optics 
\cite{Dousse:2010uq,Abbarchi:2013fk,Gerace:2009kx,hamel2015,PhysRevB.86.045315}. The high optical nonlinearities combined with tight light confinement open up new routes for the study of light-matter 
interaction in non-equilibrium driven dissipative systems. Even at the semiclassical level, already two coupled cavities --or photonic dimers-- may display a wide range of rich nonlinear dynamical phenomena 
such as instabilities and bifurcations \cite{hamel2015,PhysRevB.77.125324,Rodriguez:2016fk}.

In this context, micro and nanolasers prove useful to investigate nonlinear dynamics in the low photon number regime. Either cavity quantum electrodynamics (cQED) microlasers or photonic band-gap materials can provide large enough spontaneous emission factors ($\beta$) such that strong nonlinearities may take place in an optical cavity with, ultimately, few photons. Large Purcell factors in the former, and the suppression of leaky modes in the latter can be tailored in order to obtain high $\beta$-factors. As a consequence, the saturation photon number, $s_{sat}\sim \beta ^{-1}$, may be significantly lowered \cite{PhysRevA.50.4318}. Micro and nanolasers may display rich nonlinear dynamical 
behaviors even at a single laser level, such as injection locking phenomena \cite{schlottmann2016injection}, self-pulsing \cite{yu2016demonstration} and superradiant giant photon bunching \cite{Jahnke:2016vn}. 
Recently, spontaneous symmetry breaking in coupled photonic crystal nanolasers has been demonstrated with $\sim 150$ intracavity photons ($\beta \sim 0.02$) \cite{hamel2015}, paving the way to explore a wealth of nonlinear dynamical phenomena in photonic dimers operating in a laser regime. These systems can also be useful for applications such as flip-flop optical memories and logic gates in active photonic circuitry \cite{liu2010}.
In this article we show, both theoretically and experimentally, the asymmetric interaction between strongly coupled modes in a photonic crystal nanolaser dimer. 
We will analyze this phenomenon in the framework of "two discrete sites" Bogatov effect, in the sense that the dynamic population grating takes the form of oscillations of the population imbalance between two optical defects at the Rabi frequency (Fig.~\ref{fig1}a). 

\section{The two-site Bogatov effect}
\begin{figure}[!t]
\centering
\includegraphics[scale=0.4]{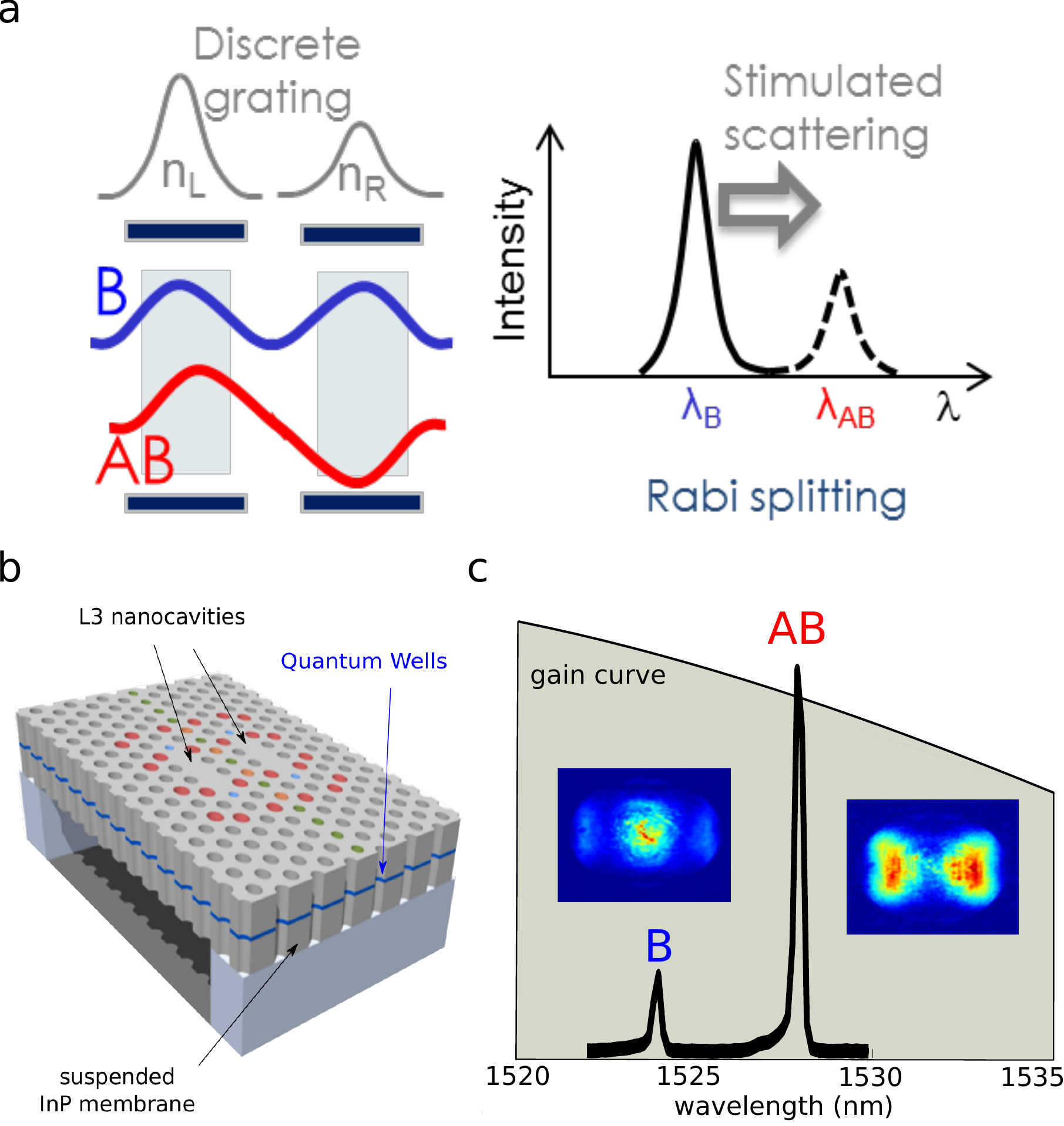}
\caption{a) Schematic representation of the eigenmodes of the coupled nanolasers and carrier population in each cavity. The dynamic population grating oscillates at the Rabi frequency and induces stimulated scattering from the blue to the red mode. b) Coupled PhC nanolasers suspended on an InP membrane. Green and orange circles: modified barrier holes. Red circles: modified holes for better beaming efficiency. c) Optical spectrum of the AB and B resonances with the corresponding far-field patterns. Grey: schematic representation of the profile of the gain curve.}  
\label{fig1}
\end{figure}

Mode competition is at the heart of multimode laser dynamics, and may occur as long as different modes share the same gain medium.  For instance, mode switching and bistability 
have been observed in standard microcavity systems, such as vertical cavity semiconductor lasers (VCSELs) \cite{Kawaguchi2010} and micropillar lasers \cite{Leymann2013,redlich2016mode}, 
and they have also been predicted in photonic crystal coupled cavities \cite{PhysRevLett.99.073902}. These mechanisms usually rely on cross-gain saturation effects. 
However, mode interaction in semiconductor cavities may also result from stimulated scattering due to a dynamic carrier population grating in the 
gain medium, which oscillates at the beat note frequency between adjacent cavity modes ($\sim10-100$ GHz). 
Such interaction, known as the "Bogatov effect", is asymmetrical in the sense that the laser mode at the 
blue side of the spectrum transfers its energy to the mode at the red side \cite{bogatov1975anomalous}. This phenomenon is a consequence of nonlinear dispersion, and it is also known to occur in free-running edge-emitters \cite{PhysRevA.69.053816,Lenstra:14}, 
in VCSELs supporting two orthogonal polarizations \cite{choquette1995}, and in vertical external cavity lasers (VECSELs) leading to a slow light regime \cite{PhysRevLett.105.223902}. 
Yet, the observation of this effect in photonic dimers remains elusive to date. 
For moderate $\beta$-factors ($\beta \sim 0.01$) as in our case, 
an accurate description of the system can still be obtained in terms of mean-field equations accounting for the dynamics of two coupled complex field amplitudes in the left (L) and right (R) nanocavities filled with a quantum well gain medium \cite{hamel2015}(see Fig.~\ref{fig1}b). The evolution of the field and carrier densities are governed by 
\begin{align}
\dot{a}_{L,R}  &=   -\kappa a_{L,R}+ \frac{\beta \gamma_{\parallel}}{2} \left(1+i\alpha\right)\left(n_{L,R}-n_0\right)a_{L,R}+  \nonumber \\
 & + \left(\gamma+iK\right)a_{R,L} +F_{a_{R,L}}(t) \label{eq:aLR}\\
\dot{n}_{L,R}  &=  p_{L,R}-\gamma_{tot} n_{L,R}-\beta \gamma_{\parallel} \left( n_{L,R}-n_0\right) |a_{L,R}|^2 + F_{n_{R,L}}(t)\label{eq:nLR}
\end{align}
where $|a|^2$ and $n$ are normalized as the photon and carrier numbers in the cavities, respectively, $\kappa$ is the cavity loss rate, $\gamma_{\parallel}$ is the two-level radiative recombination rate, $\alpha$ the Henry factor, 
$n_0$ the carrier number at transparency, $p_{L,R}$ the pump rate and $\gamma_{tot}$ is the total carrier recombination rate. The complex inter-cavity coupling constant quantifies frequency ($K$) and loss ($\gamma$) splitting as a result of the evanescent coupling, and $F_{a}$ are Langevin noise terms accounting for spontaneous emission with rate  $R_{sp}= \beta B n_{L,R}^2/V_a$ where $B$ is the bimolecular radiative recombination rate and $V_a$ the volume of the active medium. The expression of $F_{n}$ is standard; it is chosen as to preserve the proper correlations between the photon and carrier numbers, see \cite{hamel2015} Supplementary Material for more details. In order to address mode interaction, we first project Eqs.~\ref{eq:aLR} and \ref{eq:nLR} on the basis of eigenmodes,  $a_B=(a_L+a_R)/\sqrt{2}$ and $a_{AB}=(a_L-a_R)/\sqrt{2}$, corresponding to bonding (B) and anti-bonding (AB) modes of the photonic dimer respectively (Fig.~\ref{fig1}a). 
Dynamic grating Bogatov effects can be described by considering a superposition of monochromatic fields, $a_{B,AB}=A_{B,AB} \exp(\pm iK t)$, which induces oscillation of the carrier number at the Rabi frequency $2K$: 
\begin{eqnarray}
\bar{n} & = & \bar{n}_0+\left( \bar{N} e^{2iK t} +c.c.\right)  \\
\Delta n& = & \Delta n_0+\left( \Delta N e^{2iK t} +c.c.\right) 
\end{eqnarray}
where $\bar{n}=(n_L+n_R)/2$ is the mean carrier number, $\Delta{n}=(n_L-n_R)/2$ is the population imbalance between the cavities, $\bar{n}_0$ and $\Delta{n}_0$ are the steady state values, and $\bar{N}$ and $\Delta{N}$ are small amplitude complex oscillations. Replacing these expressions into Eqs.~\ref{eq:aLR} and \ref{eq:nLR}, and neglecting spontaneous emission for the moment, we obtain:
\begin{equation}
\Delta N=-\frac{\beta \gamma_\parallel}{4 \gamma_{tot}}\cdot \frac{(\bar{n}_0 -n_0)A_BA_{AB}^*}{1+\frac{\beta \gamma_\parallel}{2 \gamma_{tot}}(|A_B|^2+|A_{AB}|)^2+\frac{2iK}{\gamma_{tot}}}
\label{grating}
\end{equation}
Equation \ref{grating} stands for the discrete grating complex amplitude, and it is at the origin of asymmetric interaction between the modes (Fig.~\ref{fig1}a). The stimulated gain originating from Eq.~\ref{grating} for the bonding ($\delta {G}_B$) and the anti-bonding ($\delta G_{AB}$) modes can be approximated, in the limit of large Rabi splitting, by
\begin{equation}
\delta G_{B,AB}= \mp \frac{(\beta \gamma_\parallel)^2}{4 K}(\bar{n}_0 -n_0)\alpha|a_{AB,B}|^2
\label{stimgain}
\end{equation}
Equations \ref{stimgain} contains three important features: i) the mode interaction is anti-symmetric, meaning that the sign is reversed when changing the direction of the interaction (scattering from $a_B$ to $a_{AB}$ or viceversa); 
ii) for $K>0$, i.e. blue-detuned bonding mode, the positive gain contribution is experienced by the red-detuned AB mode; and iii) this effect is proportional to the $\alpha$-factor, thus revealing the nonlinear index effect as the main mechanism for the stimulated scattering. Note that saturation term in Eq.~\ref{grating} is negligible provided that $2K/\gamma_{tot}\gg s/2s_{sat}$, where $s$ is total photon number, and $s_{sat}=\gamma_{tot}/\beta\gamma_{\parallel}$. In our case this approximation is justified since $2K/\gamma_{tot} \sim 100$ and $s$ is few times $s_{sat}$ in a normal laser operation regime. 

\section{Experimental Results}
\subsection{Coupled Photonic-Crystal nanolasers}
In order to experimentally realize the discrete Bogatov effect, we have designed and fabricated two evanescently-coupled active photonic crystal (PhC) L3 cavities (three holes missing in the $\Gamma$K direction of a triangular lattice) in a semiconductor free standing membrane (Fig.~\ref{fig1}b). The size of surrounding holes has been modified to both increase beaming efficiency (band-folding technique) and control the inter-cavity coupling strength (barrier engineering). 
Details on the PhC structure can be found in Refs.~\cite{hamel2015} and \cite{haddadi2014}. 
Both single and coupled cavities have been etched in InP membranes containing InGaAs/InGaAsP quantum wells. The measured Q-factor at transparency is $Q\sim 4300$ ($\tau=\kappa^{-1}\approx 7 \, \mathrm{ps}$, where $\tau$ is the photon lifetime in the cavity) and the spontaneous emission factor ($\beta$) is $\sim$ 0.02 for coupled cavities. The linewidth enhancement factor is $\alpha = 7$.  Two modes are observed in the coupled-cavity system: the B mode for a symmetric superposition of the single cavity modes, and the AB mode for an anti-symmetric superposition, which can be clearly identified in the far field images (Fig.~\ref{fig1}c).
\begin{figure}[!t]
\includegraphics[scale=0.2]{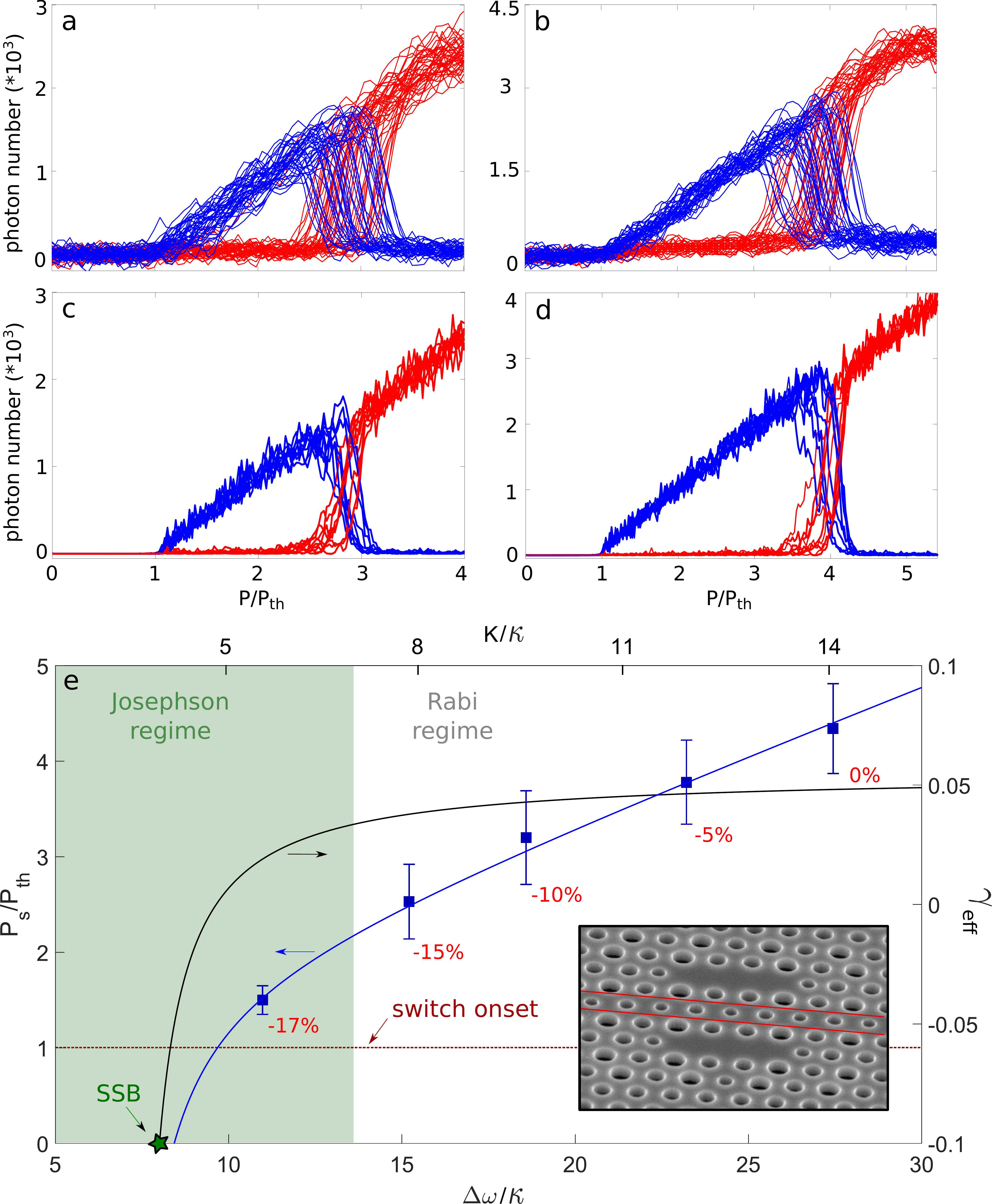}
\caption{a,b) Superposition of 40 experimental LL-curves in quasi-CW pumping showing the mode switching for K $\sim$ 8$\kappa$ (a) and K $\sim$ 12$\kappa$ (b). c,d) Superposition of 10 numerical LL-curves for K = 8$\kappa$ (c) and K = 12$\kappa$ (d). e) Black line: effective loss split parameter $\gamma_{eff}$. Markers: pump power at the switching point normalized by pump power at threshold plotted versus normalized mode splitting ($\Delta \omega$/$\kappa$, bottom axis) and Rabi frequency (K/$\kappa$, top axis). The corresponding wavelength splittings from the leftmost to the rightmost points are: 1.92, 2.66, 3.25, 4.06 and 4.8 nm. Blue line: fit of the experimental data using Eq.~\ref{Ps} and $\gamma_{eff}$ (see text). The star indicates the operation point of the Spontaneous Symmetry Breaking (SSB) transition reported in \cite{hamel2015}. The red numbers represent the modification of the holes in the central row of the PhC structure in order to tune the amount of evanescent coupling between the nanocavities. The inset shows a structure where central holes diameters were decreased by 15\%. }  
\label{switch-quasiCW}
\end{figure}
The effective refractive index of the symmetric mode of the underlying W1 waveguide (one row of holes omitted in the $\gamma$K direction of the triangular lattice) is smaller than the index of the anti-symmetric mode. Hence the B mode is blue-detuned with respect to the AB mode \cite{haddadi2014}. 
Our strategy to observe Bogatov mode interaction is to induce a mode switching by stabilizing the system close to the laser threshold at the blue most mode (B). For this, the PhC lattice parameter is tuned such that the optical resonances lie on the red slope of the gain curve (Fig.~\ref{fig1}c) (maximum at $\lambda_0$ $\sim$ 1510 nm and FWHM $\sim 63$ nm).  As a result, the B mode has larger enough gain and, even though the AB mode has lower optical losses \cite{hamel2015}, the B mode experiences larger net gain. Therefore, the blue-detuned mode (B) is the lasing one close to laser threshold. 

\subsection{Mode switching}
The coupled nanolasers are optically pumped at room temperature. We use optical periodic (50 kHz-repetition rate) and incoherent ($\lambda=800$ nm) pumping consisting either of a ramp for the acquisition of the output vs. input power (LL-curve) or short pulses ($\sim 100$ ps-pulse duration) to perform statistics on the output pulse energies. In both cases, the pump spot is located at the center of the coupled cavity system. The emission is collected with a N.A.=0.95 microscope objective, and its back focal plane is imaged through a lens to obtain the far-field 
pattern (Fig.~\ref{fig1}c). Two single mode fibers coupled to microscope objectives are used to spatially select two regions of the far field: the center corresponds to B-mode intensity, and one of the lateral lobes to the AB-mode. 
These optical signals are sent to two identical low noise (200 fW$/ \sqrt{Hz}$), 660 MHz-bandwidth avalanche photodiode (APD) detectors. This is similar to the experimental technique of Ref. \cite{hamel2015}, but now the 
two detectors monitor intensity fluctuations from modes rather than individual cavities.

The optical Rabi frequency is related to the laser mode splitting $\Delta \omega$ as $K=\Delta \omega /2 +\alpha \gamma$ \cite{hamel2015}. In Fig.~\ref{switch-quasiCW}a,b we show the simultaneous output intensities of both modes as a function of the instantaneous pump power for two different Rabi mode splittings ($K=8\kappa$ and $K=12\kappa$). 
This plot consists of a superposition of 40 sequences of quasi-CW pumping. It can be clearly observed that, 
at threshold, lasing takes place on the B-mode, demonstrating that it has higher net gain as expected. In Fig.~\ref{switch-quasiCW}a the B mode loses stability and turns off at P $\approx$ 2.5P$_{th}$, while the AB mode turns on right after. 
This reveals energy transfer from the blue-detuned to the red-detuned mode. In addition we observe that, from sequence to sequence, the instability threshold fluctuates in a range 2.1 $\leqslant$ P/P$_{th}$ $\leqslant$ 2.9. 
Once the bifurcation point is crossed, the system remains stable on the AB-mode up to large values of the pumping. A similar behavior is observed for larger coupling strength (Fig.~\ref{switch-quasiCW}b), and in this case the switching point occurs at a larger pumping level, P $\sim$ 3.8P$_{th}$. Note that this is consistent with the dependence of the stimulated gain on the cavity coupling of Eq.~\ref{stimgain}, i.e. $\delta G_{AB}\sim K^{-1}$. 
Next, we explore the full range of experimentally available coupling levels by extensively exploiting the barrier engineering technique. To this aim we have fabricated a number of different coupled cavity systems 
with varying evanescent coupling. 
Laser emission experiments on these cavities have been carried out subsequently, and the results are depicted in Fig.~\ref{switch-quasiCW}e. 
Here, the pump power at the switching point is plotted as a function of the normalized frequency splitting of the laser modes. 
\begin{figure}[!t]
\centering
\includegraphics[scale=0.17]{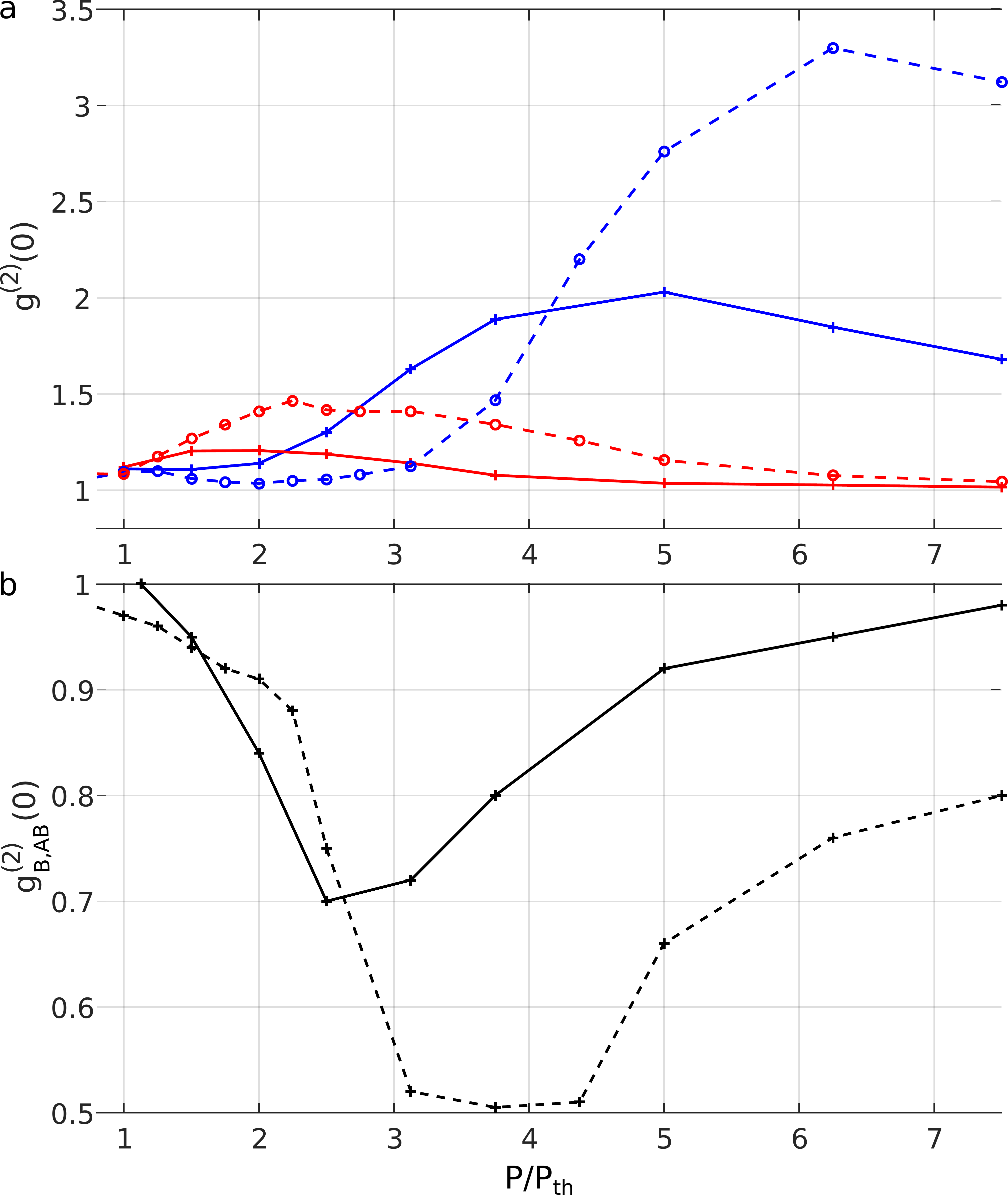}
\caption{a) Second order auto-correlation function for AB (red) and B (blue) as a function of normalized pump power. solid line: K = 8$\kappa$, dashed line: K = 12$\kappa$. b) Second order cross-correlation function as a function of normalized pump power for K = 8$\kappa$ (solid line) and K = 12$\kappa$ (dashed line).}  
\label{g2_exp}
\end{figure}

\subsection{Statistics of mode fluctuations}
Our last experimental study deals with the statistics of the laser emission. Since a few hundreds of photons fill the cavities (in our case $s_{sat} \approx 150$), stochastic fluctuations play a fundamental role, which is 
typical in nanolasers. We have investigated the consequences of this in the context of mode switching by performing a statistical study of the mode fluctuations. 
With this aim we have implemented a short-pulse pumping scheme developed in \cite{marconi2016} which allows us to obtain second order auto and cross-correlation functions of the emitted photon number $m$:
\begin{eqnarray}
g^{(2)}(0) & \approx & 1 + \frac{(\Delta m)^2}{{\langle m \rangle}^2} \label{g2}\\
g^{(2)}_{B,AB}(0) & = & \frac{\langle m_{B} m_{AB} \rangle}{\langle m_{B} \rangle \langle m_{AB} \rangle}  \label{g2cross}
\end{eqnarray}
where $\Delta m$ is the standard deviation. 
Note that Eq.~\ref{g2} holds in the large photon number limit \cite{loudon}. 

Figure \ref{g2_exp} depicts self and cross second order correlations functions obtained from time traces containing $10^4$ pulses, with $K=8\,\kappa$ (solid line) and $K=12\,\kappa$ (dashed line). 
Before the switch, lasing occurs in the B-mode and thus $g_B^{(2)}(0) \sim $ 1 while the AB mode develops super-Poissonian fluctuations. We point out that in this close-to-threshold pump region --highly influenced by noise--, 
the actual $g_{AB}^{(2)}(0)$ might be larger compared to the observed one, which is limited by the resolution of our experimental technique. 
A decrease of $g_{B,AB}^{(2)}(0)$ is visible, which is the signature of a growing anti-correlation of the mode fluctuations. 
The mode switching occurs as $g_{B}^{(2)}(0) \approx g_{AB}^{(2)}(0)$ which takes place at $P_s\approx 2.2P_{th}$ with a value of $g_{switch}^{(2)}(0)\approx 1.2$ for $K=8\,\kappa$, and $P_s\approx 3.5 P_{th}$ with a value of $g_{switch}^{(2)}(0)\approx 1.4$ for $K=12\,\kappa$. Note that at the switching points the modes become indistinguishable in terms of the second moments of their energy fluctuations. 
Furthermore, for $K=12\,\kappa$, a "soft" region for 3 $\leqslant$ P/P$_{th}$ $\leqslant$ 4.3 is characterized by a plateau 
in $g_{B,AB}^{(2)}(0) \approx 0.5$, which corresponds to both modes present in the optical spectrum. This region also exists when the coupling is smaller, but the plateau has a larger cross-correlation, 
$g_{B,AB}^{(2)}(0) \approx 0.7$. This could be due to a stronger effect of spontaneous emission noise for small K as a result of a closer proximity of the switching point to the laser threshold. 

The experimental results can be compared with the full solutions of the coupled field equations in presence of spontaneous emission, that we modeled as Langevin forces. 
Figures~\ref{switch-quasiCW}c,d) show simulated time traces for two coupling parameters $K=8\,\kappa$ and $K=12\,\kappa$. Note that the energy transfer in the form of a switch mechanism is well reproduced by the model.  
The jitter of the mode switching, in turn, is larger in the experiment, which could be explained as a consequence of other sources of noise (e.g. mechanical vibrations and thermal effects).

\section{Theoretical analysis}

We now theoretically investigate the nonlinear dynamical origin of the mode switching and explain the dependence of the switch-power upon the coupling parameter observed in Fig.~\ref{switch-quasiCW}e. 
To this aim, we have performed an analytical reduction of the deterministic part of Eqs.~\ref{eq:aLR} and \ref{eq:nLR} ($F_a=F_N=0$) to a one-dimensional (1D) dynamical system able to capture the essence of the mode switching dynamics. 
By exploiting the specific scaling of the parameters and
of the governing time scales, it is possible to reduce the time evolution of the system 
to the dynamics of the fractional photon population imbalance between
the two modes, $h=\left(|a_B|^2-|a_{AB}|^2\right)/\left(|a_B|^2+|a_{AB}|^2\right)$. 
In this framework, the B and AB solutions are represented by the two extreme values $h=\pm1$. 
Under the assumption of large Rabi splitting and small population imbalance it is possible to show that the dynamics of $h$ is approximately ruled by the following 1D differential equation:
\begin{equation}
\dot{h}  =  -\frac{dU(h)}{dh} \label{eq:zztop}
\end{equation}
where $U(h)=-\gamma_h (h-h^3/3)$ is an effective 1D potential. 
The factor $\gamma_h$ is a 1D nonlinear optical loss splitting rate which reads: 
\begin{equation}
\gamma_{h} =  2\gamma-\frac{\alpha \gamma_{tot}}{2K} \left(P-1\right)\label{eq:gamz}
\end{equation}
\begin{figure}[!t]
\includegraphics[bb=20bp 0bp 630bp 451bp,clip,width=1\columnwidth]{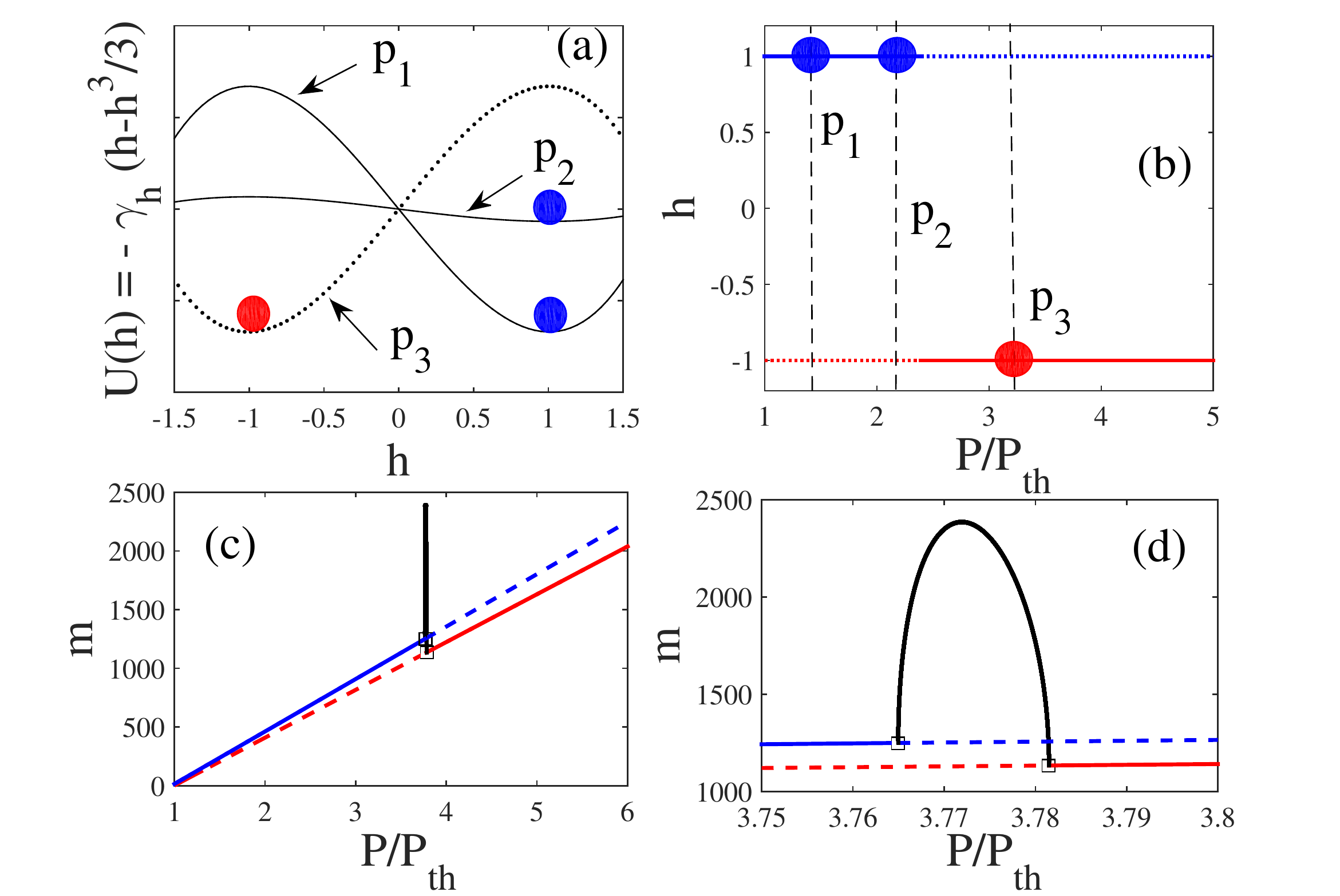}
\caption{(a) Sketch representing the exchange of stability between the B (in blue) and the AB mode (in red) depicted in (b). The values of $\gamma_h$ corresponding to the three values of the normalized pump power $p_{1,2,3}=P_{1,2,3}/P_{th}$ are $\gamma_h=1,\,0.1$ and $-1$.  (c) Bifurcation diagrams for the photon number in a single cavity in the B and AB modes as a function of pump power ($K=12\kappa$). A switching occurs from the B to the AB mode through a Hopf bifurcation at $P=3.765\,P_{th}$. (d) Blow-up around the bifurcation points and the narrow periodic bridge solution. Full and dotted lines correspond to stable and unstable solutions, respectively.}  
\label{theo}
\end{figure}
Equations \ref{eq:zztop} and \ref{eq:gamz} contain the essence of the mode switching phenomenon induced by the Bogatov effect in the coupled cavity system. One of the fixed points $h=\pm1$ is stable (node) and the other one unstable (saddle), and depending on the sign of $\gamma_{h}$ the stable one will be bonding ($h=1$) or anti-bonding ($h=-1$).
This is illustrated in Fig.~\ref{theo}a, where the potential $U(h)$ is represented for three different pumping parameters: before ($\gamma_h>0$) and after ($\gamma_h<0$) the switching point, and for an intermediate pump level ($\gamma_h=0.1$). 
The mode switching depicted in Fig.~\ref{theo}b is thus represented by the change of sign of $\gamma_{h}$. Setting $\gamma_{h}=0$ in Eq.~\ref{eq:gamz} allows us to find the locus of the mode switching pump power $P_s$ as a function of
the imaginary ($K$) and real ($\gamma$) parts of the coupling parameter:
\begin{equation}
P_s=\left( 1+\frac{4\gamma K}{\alpha \kappa \gamma_{tot}} \right)\cdot \frac{2\kappa}{\beta \gamma_{\parallel} N_0}+1 \label{Ps}
\end{equation}
When using Eq.~\ref{Ps} to fit the experimental results, one has to take into account the gain difference between B and AB modes, since they are spectrally located on the slope of the gain function 
(see Fig.~\ref{fig1}c). A natural way to include this effect in the model is to replace $\gamma$ by an effective $\gamma_{eff}$ parameter which accounts for net optical losses (loss split minus gain difference). 
Hence, the effective loss split parameter depends on mode spacing, $\gamma_{eff} (\Delta \omega)$. We consider the following phenomenological saturable function 
$\gamma_{eff} (\Delta \omega)=\gamma_1+(\gamma_0-\gamma_1)/[1+(\Delta \omega-\Delta \omega_0)/\Delta \omega_{sat}]$, where $\gamma_0=\gamma_{eff}(\Delta \omega_0)$ is the "zero-detuning" loss split, 
$\gamma_1$ is the large $\Delta \omega$ limit, and $\Delta \omega_{sat}$ is a saturation parameter. We fix $\gamma_0=-0.1$ for $\Delta \omega=8\kappa$ (or $K=3.3 \kappa$), for consistency with the 
operation point of the symmetry breaking (SSB) transition, whose parameters have been determined in \cite{hamel2015}. 
Two effects are captured by $\gamma_{eff}$($\Delta \omega$): a first-order linear increase for $\Delta \omega$ $\approx$ $\Delta \omega_0$ which comes from the non-zero slope of the gain curve (Fig.~\ref{fig1}c), and higher-order saturation terms. The latter can be attributed to saturation effects not taken into account in the original model, such as spectral hole burning \cite{agrawal}.  
Our expression of $\gamma_{eff} (\Delta \omega)$ is inserted into Eq.~\ref{Ps} to fit the experimental data, 
and the result is plotted in Fig.~\ref{switch-quasiCW}e. It is interesting to note that two regimes can be identified: the Rabi regime, where the nonlinear laser frequency shift is smaller than the Rabi splitting, i.e. $K/\kappa>\alpha$, 
and the Josephson regime for weaker Rabi splitting, i.e. $K/\kappa<\alpha$. The SSB transition lies within the Josephson region (see Fig.~\ref{switch-quasiCW}e), where ultra fast ($\sim 150$ GHz) oscillations have been predicted 
to emerge through secondary (Hopf) bifurcations \cite{hamel2015}. 
In this regime, $\gamma_{eff}$ is rapidly increasing and Bogatov mode interaction can be observed as well (see point at $\Delta \omega=11\kappa$). Indeed, for $\Delta \omega=11\kappa$, also SSB can be observed after the mode switching instability. For increasing $\Delta \omega$ values we enter into the Rabi regime, where $\gamma_{eff}$ strongly saturates to $\gamma_{eff}\approx \gamma_1=0.055$. In this regime of 
nearly constant $\gamma_{eff}$, the linear dependence of the switch power upon the coupling parameter K predicted in Eq.~\ref{Ps} is established. 

Within this simple description, the switching point takes place at a single point $P_s$ since the switching mechanism is represented by an exchange of stability between the two solutions. First, the presence of spontaneous emission modifies this  picture and induces a "transition region" in the pump parameter with a nonzero measure. Second, the single bifurcation point is a consequence of the large $K$ approximation used to derive Eq.~\ref{eq:zztop}.  Relaxing this hypothesis, i.e. for finite K-values, the switching mechanism becomes more complex: a non-trivial connection of B and AB solutions in phase space comes up as the pump parameter is varied. This question can be elucidated by computing a bifurcation diagram for the full deterministic dynamical system of Eqs.~\ref{eq:aLR} and \ref{eq:nLR}, shown in Fig.~\ref{theo}c. It can be observed that the stable B-mode undergoes a supercritical Hopf bifurcation at P $\sim$ 3.765 P$_{th}$. In other words, there is a switching zone (see the inset of Fig.~\ref{theo}c) where the only stable solution is a limit cycle, which manifests itself as a cavity beating at the Rabi frequency. This limit cycle, in presence of noise, is the nonlinear dynamical representation of the soft transition region from the blue-detuned to the red-detuned mode where the coupled nanolasers operate in a dual-frequency regime. Such a fast nonlinear beating dynamically delocalizes photons in the coupled cavity system. As a result strong anti-correlated intensity fluctuations for B and AB modes are expected, which is consistent with the observed minimum of the cross-correlation function at the switching point (Fig.~\ref{g2_exp}). Hence, we conclude that the observation of a dip in the mode cross-correlation is the statistical consequence of a dual frequency operation region given by an ultrafast ($\sim0.5$ THz) noisy limit cycle rather than, for instance, bi-stable mode switching scenarios previously reported for bi-modal microlasers \cite{Leymann2013,redlich2016mode}.

\section{Conclusion}
In conclusion, we have revealed for the first time asymmetric energy transfer between eigenmodes in a semiconductor photonic dimer. We have experimentally shown that a mode switching occurs from the blue-detuned (bonding) to the red-detuned (anti-bonding) modes as the pump power is increased. We have identified the basic mechanism underlying such mode interaction as a "discrete Bogatov effect", i.e. the asymmetric stimulated light scattering induced by two-site population oscillations at the optical Rabi frequency. The predicted scaling of the onset of switching with respect to the coupling parameter in a mean-field model is shown to be in good agreement with the experimental results for different coupling strengths. We claim that this phenomenon is generic in semiconductor coupled cavities, and dominates against cross-gain saturation effects, therefore it 
could be used to model nonlinear mode coupling in probabilistic theories such as bimodal birth-death models in micro/nano-laser photonic dimers. Furthermore, it should also be scalable to N-coupled cavities, such that energy can be expected to flow from the blue-most to the red-most hybrid modes of the photonic molecule with the increase of pump power.

\section*{Acknowledgments}
The authors acknowledge I. Sagnes and G. Beaudoin for the fabrication of the samples. This work is supported by a public grant overseen by the French National Research Agency (ANR) as part of the "Investissements d'Avenir" program (Labex NanoSaclay, reference: ANR-10-LABX-0035) and funding from the ANR project OPTIROC. J.J. acknowledges financial support from the Ram\'on y Cajal fellowship and project COMBINA (TEC2015-65212-C3-3-P). 


\end{document}